\documentclass[12pt,a4paper]{article}
\usepackage{amsfonts,latexsym}
\usepackage{amsmath,amssymb}
\usepackage{graphicx,color}

\renewcommand{\thefootnote}{\fnsymbol{footnote}}

\begin{document}

\vspace{12mm}

\begin{center}
{{{\Large {\bf Thermodynamic and tachyonic instability for asymptotically flat black holes }}}}\\[10mm]

{Yun Soo Myung$^{1}$\footnote{e-mail address: ysmyung@inje.ac.kr}, De-Cheng Zou$^{2}$\footnote{e-mail address: zoudecheng789@hotmail.com}
and Meng-Yun Lai$^2$\footnote{mengyunlai@jxnu.edu.cn;}}\\[8mm]

{${}^1$Institute of Basic Sciences and Department  of Computer Simulation,  Inje University, Gimhae 50834, Korea\\[0pt]}

{${}^2$College of Physics and Communication Electronics, Jiangxi Normal University, Nanchang 330022, China\\[0pt]}
\end{center}

\vspace{2mm}

\begin{abstract}
It was confirmed that the negative modes of the Euclidean section for  asymptotically flat black holes reveal the thermodynamic instability of these black holes in the grand canonical ensemble (GCE). These include Schwarzschild, Reissner-Nordstr\"{o}m, Kerr, and Kerr-Newman black holes.
In this work, we develop the relation between thermodynamic instability in the GCE and tachyonic instability for asymptotically flat black holes, where the latter is the onset for obtaining black holes with scalar hair.
This implies that the tachyonic instability of black holes when introducing scalar coupling to the Gauss-Bonnet term or Maxwell term  reflects thermodynamic instability of these black holes in the GCE. We go on further to consider the Schwarzschild-AdS black hole. 
\end{abstract}
\vspace{5mm}

\newpage
\renewcommand{\thefootnote}{\arabic{footnote}}
\setcounter{footnote}{0}


\section{Introduction}
Gross, Perry, and Yaffe (GPY)~\cite{Gross:1982cv} have found that a nonconformal negative mode of Euclidean section of Schwarzschild black hole (Schwarzschild instanton)
is related to the negative specific heat, implying  local thermodynamic instability in the canonical ensemble (CE).
A negative mode  of Reissner-Nordstr\"{o}m (RN) instanton disappeared when the specific heat at constant charge becomes positive in the CE~\cite{Monteiro:2008wr}, indicating  thermodynamic instability in the CE. However, it was argued that there is always one negative eigenvalue and the grand canonical ensemble (GCE) is thermodynamically  unstable because the specific heat at constant charge and isothermal permittivity  always have opposite signs.  Also, it was shown that  a negative mode of Kerr quasi-instanton persists when the specific heat at constant angular momentum is positive and the isothermal moment of inertia is negative. This reveals thermodynamic instability in the GCE because the thermodynamic  stability of the GCE further requires the positivity of the isothermal moment of inertia~\cite{Monteiro:2009tc}. The Kerr-Newman (KN) black hole seems to be thermodynamically  unstable in GCE~\cite{Ruppeiner:2007hr,Ruppeiner:2008kd} even though the negative mode is not yet known.
 The above indicates a close relationship between  negative mode of black hole instanton and thermodynamic instability of black hole in the GCE.

At this stage, we would like to mention the dynamical stability of black holes, which confirms the existence of these black holes in curved spacetimes~\cite{chan}.
If a solution of the black hole is dynamically unstable, it is not considered as a true black hole.
For Schwarzschild black hole, the Regge-Wheeler prescription works to indicate no dynamical instability~\cite{Regge:1957td,Zerilli:1970se}.
It was shown that the  RN black hole is stable against the tensor-vector perturbations~\cite{Zerilli:1974ai,Moncrief:1974ng}.
The Kerr black hole is stable against the gravitational perturbations~\cite{Press:1973zz}.
The slowly rotating KN black hole is stable~\cite{Pani:2013ija,Pani:2013wsa} and the nonlinear stability of near-extremal KN black hole is investigated to be stable~\cite{Zilhao:2014wqa}.
Therefore, the analysis shows that the dynamical stability of these black holes has nothing to do with thermodynamic instability in the GCE.

On the other hand, recently, there is a significant progress on obtaining black holes with scalar hair via spontaneous scalarization.
In this case,  the tachyonic instability of  black holes is regarded as  the hallmark for emerging  scalarized black holes when introducing scalar couplings to the Gauss-Bonnet term ($R^2_{\rm GB}$) for Schwarzschild black hole~\cite{Doneva:2017bvd,Silva:2017uqg,Antoniou:2017acq} and  Kerr black hole~\cite{Cunha:2019dwb} or Maxwell term ($F^2$) for RN black hole~\cite{Herdeiro:2018wub} and  KN black hole~\cite{Lai:2022ppn}. Intuitively, the tachyonic instability is realized when the negative region of scalar potential is developed near the horizon by including a negative mass squared $\mu^2=-2\alpha\bar{ R}^2_{\rm GB}$  or $\mu^2=\alpha \bar{F}^2/2$ with a positive coupling parameter $\alpha$. It is worth noting that  this instability is possible to occur for any thermodynamic state with negative or positive specific heat,  allowing for the GCE.

Now, one has to understand what is an origin of tachyonic instability for bald  black hole without scalar hair. The Euclidean negative mode found by GPY~\cite{Gross:1982cv} for the Schwarzschild black hole is related to the Gregory-Laflamme (GL) instability of the five-dimensional black string~\cite{Gregory:1993vy}.
Actually, the (thermodynamical) Euclidean negative mode of the Schwarzschild black hole [$(\Delta_{\rm L}h)_{ab}^{4D}=\lambda_{\rm GPY} h^{4D}_{ab}$] corresponds to the classical GL instability of the black string $(\Delta_{\rm L}h)_{ab}^{4D}$ $=-k^2_{\rm GL} h^{4D}_{ab}$, leading to $k^2_{\rm GL}=-\lambda_{\rm GPY}$. Here, one has $\lambda_{\rm GPY}=-0.7677$. When translating this tensor Lichnerowicz equation into the Schr\"{o}dinger-type equation, one finds the Zerilli-type potential for $s$-mode which contains a negative region outside the event horizon~\cite{Lu:2017kzi}. This captures the GL instability for obtaining  non-Schwarzschild black hole (black hole with Ricci-tensor hair)~\cite{Lu:2015cqa}. In case of  tachyonic instability for obtaining black hole with scalar hair, one usually use the linearized scalar potential  which shows a negative region  near the horizon, instead of the Zerilli-type potential~\cite{Myung:2018iyq,Myung:2018vug}. Two instabilities represents the onset for black hole with hairs: GL instability for Ricci-tensor hair and tachyonic instability for scalar hair.

In this work,  we wish to develop a close relation between thermodynamic instability in GCE (but not CE) and tachyonic instability for asymptotically flat black holes.
For this purpose, we compute all thermodynamic quantities in the GCE  and threshold curves for the onset of spontaneous scalarization.
If this relation  is confirmed, it  implies that  tachyonic instability of asymptotically flat black holes  reflects thermodynamic instability of these black holes in the GCE.
This is true for Schwarzschild-AdS (SAdS) black hole. 

\section{Static black holes }
\subsection{Thermodynamic instability in GCE}

The specific  heat  of Schwarzschild (S) black hole with mass $M$ is always  negative [see (Left) Fig. 1]
\begin{equation}
C^{\rm S}=\Big(\frac{\partial M}{\partial T}\Big)=-8\pi M^2,\label{h-sch}
\end{equation}
which indicates a typical sign for thermodynamic instability in the CE.
Also, the Rayleigh-Ritz functional for a nonconformal mode of Schwarzschild instanton takes the form~\cite{Gross:1982cv,Monteiro:2009tc}
\begin{equation}
{\cal I}^{\rm S}=\frac{\lambda_{\rm GPY}}{r_+^2}=-\frac{0.19}{M^2},
\end{equation}
which is always  negative.
It shows that the canonical ensemble breaks down for this black hole but its micro-canonical ensemble is well-defined.

On the other hand, the thermodynamic stability condition for the RN black hole with mass $M$ and charge $Q$ in the CE is given by a positivity of the specific heat at constant charge
\begin{equation}
C^{\rm RN}_Q=-\frac{2\pi r_+^2(r_+-r_-)}{r_+-3r_-}=-\frac{2\pi M^2(1+\sqrt{1-q^2})^2\sqrt{1-q^2}}{2\sqrt{1-q^2}-1}, \label{heat-rn}
\end{equation}
where the outer/inner horizons take the form  $r_\pm=M(1\pm \sqrt{1-q^2})$ with $q=Q/M$.
The positive region occurs for $\frac{\sqrt{3}}{2}<q<1$, which is the range ($q> \frac{\sqrt{3}}{2}$) for the disappearance of a negative mode
\begin{equation}
L^{\rm RN}=-\frac{2\sqrt{1-q^2}-1}{64M^2(1-q^2)^2(2\sqrt{1-q^2}+1)^2} \label{lam-e}
\end{equation}
in the partition function [see (Left) Fig. 2]~\cite{Monteiro:2008wr}.
 \begin{figure*}[t!]
   \centering
  \centering
  \includegraphics{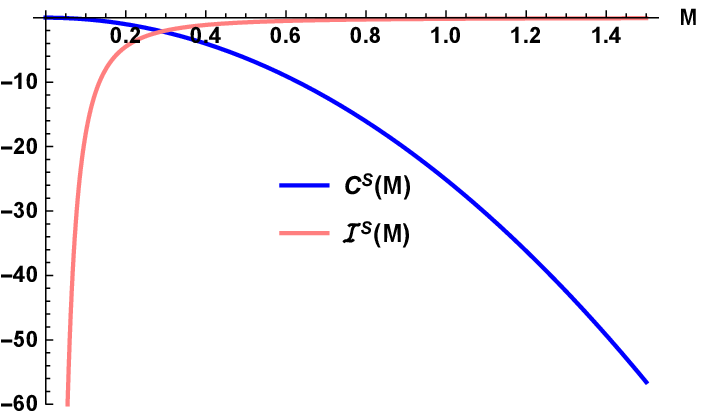}
   \hfill%
  \includegraphics{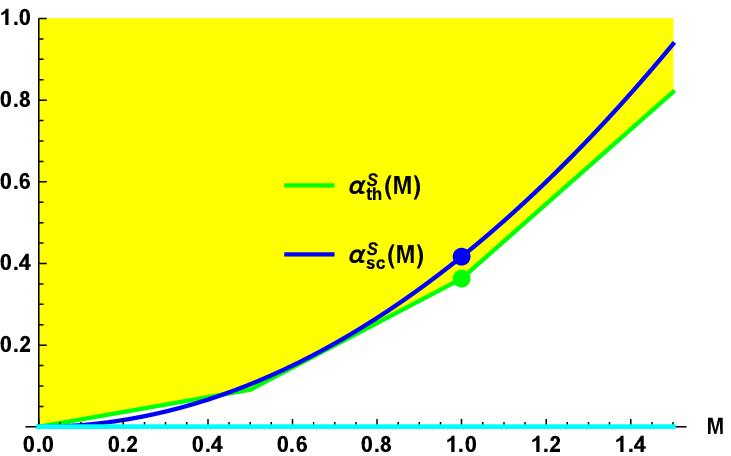}
\caption{(Left) Specific heat $C^{\rm S}(M)$ and  negative mode ${\cal I}^{\rm S}(M)$  for Schwarzschild black hole. (Right) Threshold curve $\alpha^{\rm S}_{\rm th}(M)$ and sufficient condition curve $\alpha^{\rm S}_{\rm sc}(M)$ of tachyonic instability for  Schwarzschild black hole are  increasing functions of $M$  when introducing the GB-scalar coupling. The upper (yellow) and lower (white) regions represent the unstable and stable regions, respectively. The green and  blue dots denote (1,0.363) and (1,0.4167), and  the cyan line shows a negative specific heat.   }
\end{figure*}
At this stage, one would expect a negative mode to persist in the grand partition function. It is not easy to compute this mode in this partition function.
 Now, let us introduce the grand canonical ensemble (GCE) by considering the Weinhold metric  whose inverse metric is given by
 \begin{equation}
 g_{W,RN}^{\mu\nu}=-\frac{\partial^2G}{\partial y_\mu \partial y_\nu},\quad y_\mu=(T,\Phi).
 \end{equation}
Here $G=M-TS-\Phi Q$ is the Gibbs free energy with $\Phi=Q/r_+$ the electric potential at the horizon.
In this case, we have
\begin{eqnarray}\label{gRN-m}
   g_{W,RN}^{\mu\nu} &=&  \left(
           \begin{array}{cc}
             \beta C_\Phi^{\rm RN} & \eta^{\rm RN}  \\
             \eta^{\rm RN}  & \epsilon_T^{\rm RN} \\
           \end{array}
         \right),
   \end{eqnarray}
  where $C_\Phi^{\rm RN}=-2\pi r_+^2$ is the specific heat at constant electric  potential,  $\eta^{\rm RN}=-\frac{4\pi\sqrt{r_+r_-}r_+^2}{r_+-r_-}$, and $\epsilon_T^{\rm RN}=\frac{r_+(r_+-3r_-)}{r_+-r_-}=-\frac{2\pi r_+^3}{C^{\rm RN}_Q}$ is the isothermal permittivity.  The matrix (\ref{gRN-m}) could be diagonalized by choosing a new basis of $(dT,d\Phi)\to (dT,d\Phi+\eta^{\rm RN} dT /\epsilon_T^{\rm RN})$ as
\begin{eqnarray}\label{gRN-n}
   \tilde{g}_{W,RN}^{\mu\nu} &=&  \left(
           \begin{array}{cc}
             \beta C_Q^{\rm RN} & 0  \\
            0  & \epsilon_T^{\rm RN} \\
           \end{array}
         \right)
   \end{eqnarray}
which states that there is always one negative eigenvalue and the GCE is thermodynamically unstable because   $C_Q^{\rm RN}$ and $\epsilon_T^{\rm RN}$ always have opposite signs [see (Left) Fig. 2].

\subsection{Tachyonic instability with scalar couplings}

We start with  the  Einstein-Gauss-Bonnet-scalar (EGBS) theory defined by~\cite{Doneva:2017bvd,Silva:2017uqg}
\begin{eqnarray}S_{\rm EGBS}=\frac{1}{16 \pi }\int d^4 x\sqrt{-g}
\Big[R-\frac{1}{2}(\partial\phi)^2+ \alpha \phi^2 R_{GB}^2\Big],
\label{egbs}
\end{eqnarray}
where $\alpha$ is a positive coupling parameter and $R_{GB}^2=R^2-4R_{\mu\nu}R^{\mu\nu}+R_{\mu\nu\rho\sigma}R^{\mu\nu\rho\sigma}$  represents the Gauss-Bonnet term.
The GR solution (Schwarzschild black hole) is obtained without scalar hair.
To study the tachyonic instability, we obtain  the linearized scalar equation from the scalar equation ($\nabla^2\phi+2\alpha R^2_{GB}\phi=0$)
\begin{equation}
\Big(\bar{\nabla}^2_{\rm S}-\mu^2_{\rm S}\Big)\delta \phi=0 \label{scalar-equa}
\end{equation}
with a negative mass squared
\begin{equation}
\mu^2_{\rm S}=-2\alpha \bar{R}^2_{\rm GB}=-\frac{96 \alpha M^2}{r^6}. \label{Smass}
\end{equation}

We adopt a scalar perturbation
\begin{equation}
\delta \phi(t,r,\theta,\varphi)=e^{-i\omega t } Y_{l m
}(\theta,\varphi)\frac{u(r)}{r}\,, \label{sep}
\end{equation}
where $Y_{l m}(\theta, \varphi)$ is spherical  harmonics with $-m\le l
\le m$. Defining a tortoise coordinate $r_*$ through $dr_*(r)=dr/(1-2M/r)$,  a radial part of the linearized
equation (\ref{scalar-equa})  leads to the Schr\"{o}dinger-type equation as
\begin{equation}
\frac{d^2u}{dr_*^2} +[\omega^2-V_{\rm S}(r)] u(r)=0,
\end{equation}
where the scalar potential $V_{\rm S}(r)$  takes the form
\begin{equation}
V_{\rm S}(r)=\Big(1-\frac{2M}{r}\Big)\Big[\frac{2M}{r^3}+\frac{l(l+1)}{r^2}-\frac{96 \alpha M^2}{r^6}\Big],
\end{equation}
where the last term induces a negative region outside the horizon.
The sufficient condition of  tachyonic instability for $l=0$-scalar mode  is given by~\cite{Dotti:2004sh}
\begin{equation}
\int^{\infty}_{-\infty}V_{\rm S}(r)dr_*=\int^\infty_{2M}\Big(\frac{2M}{r^3}-\frac{96 \alpha M^2}{r^6}\Big)dr<0
\end{equation}
which leads  to a condition
\begin{equation}
\alpha \ge \alpha_{\rm sc}^{\rm S}=\frac{5}{12} M^2=0.4167M^2.
\end{equation}
The sufficient condition  curve $\alpha_{\rm sc}^{\rm S}(M)$ shown in (Right) Fig. 1 is an increasing function of $M$.
However, the threshold condition for  tachyonic instability was found for $M=1$ as~\cite{Myung:2018iyq,Myung:2021fzo}
\begin{equation}
\alpha \ge \alpha_{\rm th}^{\rm S}=0.363
\end{equation}
when solving Eq.(\ref{scalar-equa}) with $\omega=i\Omega$ numerically.
\begin{figure*}[t!]
   \centering
  \centering
  \includegraphics{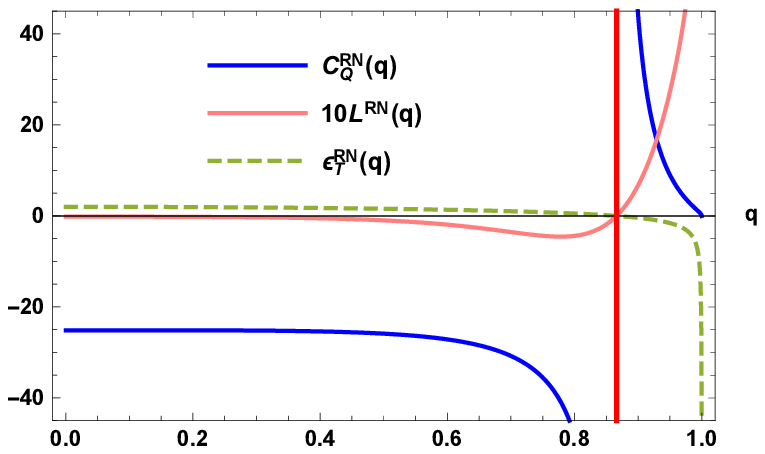}
   \hfill%
  \includegraphics{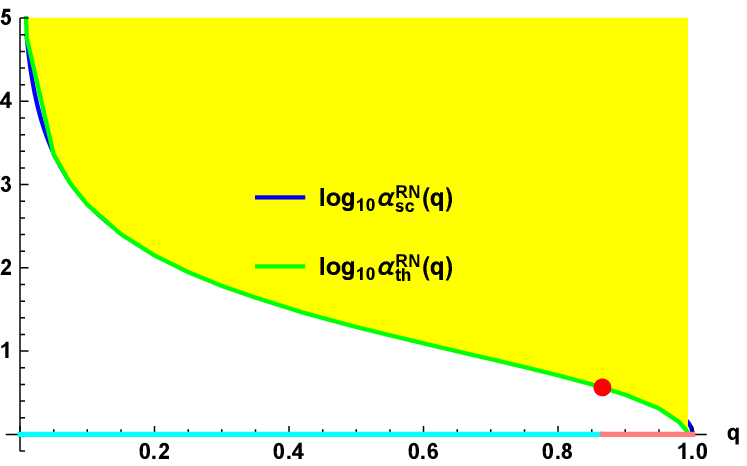}
\caption{(Left) Specific heat $C_Q^{\rm RN}(q)$, negative mode $10L^{\rm RN}(q)$, and isothermal permittivity $\epsilon_T^{\rm RN}(q)$ for RN black hole. A red line represents the diverging specific heat at $a=\sqrt{3}/2=0.866$. (Right) Threshold curve of tachyonic instability for  RN black hole is $\log_{10}\alpha^{\rm RN}_{\rm th}(q)\simeq \log_{10}\alpha^{\rm RN}_{\rm sc}(q)$ when introducing the Maxwell-scalar coupling. A red dot represents the diverging specific heat at $q=0.866$ where  cyan (pink) lines denote negative  (positive) specific heats.  }
\end{figure*}

We are in a position to introduce the Einstein-Maxwell-scalar (EMS)  theory  given by ~\cite{Herdeiro:2018wub}
\begin{equation}
S_{\rm EMS}=\frac{1}{16 \pi}\int d^4 x\sqrt{-g}\Big[ R-2(\partial\phi)^2-(1+\alpha \phi^2) F^2\Big],\label{Action1}
\end{equation}
We obtain  the scalar  equation
\begin{equation}
\nabla^2 \phi-\frac{\alpha}{2}F^2 \phi=0 \label{s-equa}.
\end{equation}
Its linearized equation is given by
\begin{equation}
\Big[\bar{\nabla}^2_{\rm RN}-\mu^2_{\rm RN}\Big] \delta\phi= 0 \label{l-eq2}
\end{equation}
with a negative mass squared
\begin{equation}
\mu^2_{\rm RN}=\frac{\alpha \bar{F}^2}{2}=-\frac{\alpha Q^2}{r^4}. \label{RNmass}
\end{equation}
Choosing a tortoise coordinate $r_*$ defined by $r_*=\int dr/f(r)$ with $f(r)=1-2M/r+Q^2/r^2$, a radial part of the linearized  equation (\ref{l-eq2}) takes the form
\begin{equation} \label{sch-2}
\frac{d^2u}{dr_*^2}+\Big[\omega^2-V_{\rm RN}(r)\Big]u(r)=0.
\end{equation}
Here the scalar potential $V_{\rm RN}(r)$ is given by
\begin{equation} \label{pot-c}
V_{\rm RN}(r)=f(r)\Big[\frac{2M}{r^3}+\frac{l(l+1)}{r^2}-\frac{2Q^2}{r^4}-\frac{\alpha Q^2}{r^4}\Big],
\end{equation}
where the last term gives rise to a negative region outside the horizon.
The sufficient condition of $l=0$-scalar mode instability is derived from $\int^{\infty}_{-\infty}V_{\rm RN}(r)dr_*<0$ as~\cite{Myung:2018vug}
\begin{equation}
\alpha \ge \alpha^{\rm {RN}}_{\rm sc}=\frac{3(1-\sqrt{1-q^2}-2q^2)}{q^2}.
\end{equation}
An actual threshold curve $\log_{10}\alpha_{\rm th}^{\rm RN}(q)[\simeq \log_{10}\alpha_{\rm sc}^{\rm RN}(q)]$ was computed numerically by solving the radial scalar equation (\ref{l-eq2}) with replacing $\omega$ by $i\Omega$  [see (Right) Fig. 2].
It is a monotonically  decreasing function of $q\in[0,1]$ and it has nothing special at a diverging specific heat ($q=0.866$).

\section{Rotating black holes}
\subsection{Thermodynamic instability in GCE}
The Rayleigh-Ritz functional ${\cal I}^{\rm K}$ for Kerr black hole was found as Eq.(50) in~\cite{Monteiro:2009tc}.
The thermodynamic stability condition for the Kerr black hole with mass $M$ and angular momentum $J$ in the CE is given by a positive value of the specific heat at constant angular momentum
\begin{equation}
C^{\rm K}_J=\frac{2\pi (r_+^2-a^2)(r_+^2+a^2)^2}{3a^4+6r_+^2a^2-r_+^4}\rightarrow_{M\to1}\frac{4\pi[3(1+\sqrt{1-a^2})-a^2(2+\sqrt{1-a^2})-a^4]}{a^4+6a^2-3}, \label{heat-k}
\end{equation}
where the outer horizon is located at $r_+=M(1+ \sqrt{1-a^2/M^2})$ with $a=J/M$ rotation parameter.
The positive region occurs for $a>\sqrt{2\sqrt{3}-3}M$, while the Rayleigh-Ritz functional ${\cal I}^{\rm K}$ is always negative and finite for $0<a<1$ [see (Left) Fig. 3]~\cite{Monteiro:2009tc}.
\begin{figure*}[t!]
   \centering
  \centering
  \includegraphics{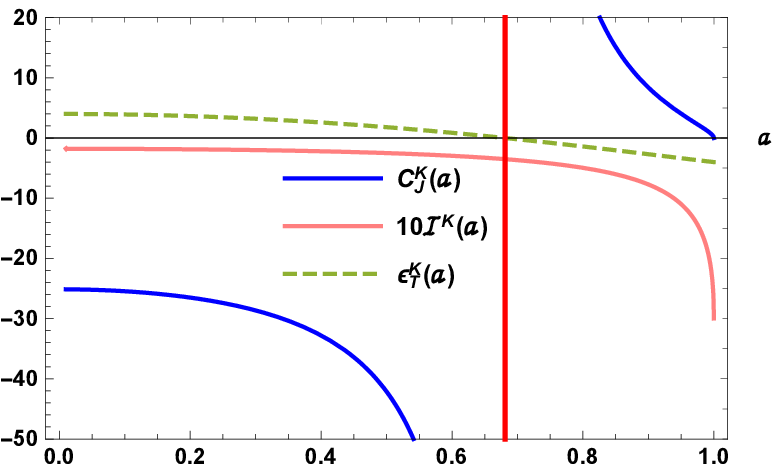}
   \hfill%
  \includegraphics{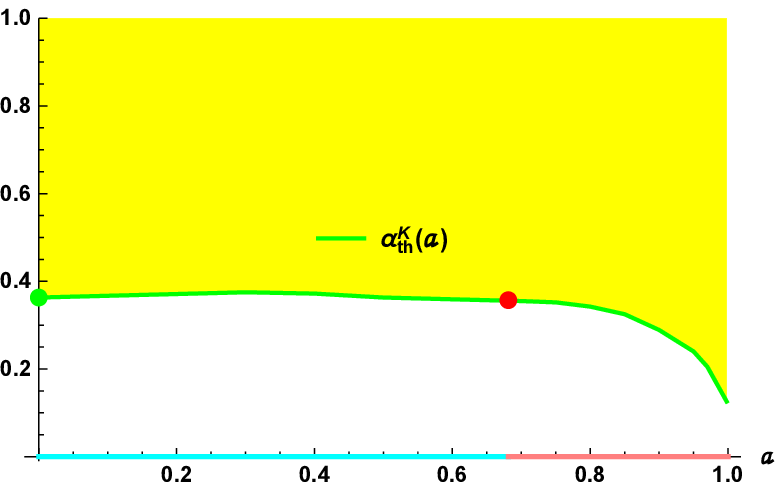}
\caption{(Left) Specific heat $C_J^{\rm K}(a)$, negative mode $10{\cal I}^{\rm K}(a)$, and isothermal permittivity $\epsilon_T^{\rm K}(a)$ for Kerr black hole. A red line represents the diverging specific heat  at $a=\sqrt{2\sqrt{3}-3}=0.681$. (Right) Threshold curve of tachyonic instability for  Kerr black hole takes the form $\alpha^{\rm K}_{\rm th}(a)$  when introducing the GB-scalar coupling. A green dot indicates the Schwarzschild black hole at $\alpha_{\rm th}^{\rm S}(M=1)=0.363$ and a red dot represents the diverging specific heat at $a=0.681$ where  cyan (pink) lines denote negative (positive) specific heats. }
\end{figure*}
Now, we wish to introduce the GCE by considering the Weinhold metric whose inverse metric is given by
 \begin{equation}
 g_{W,K}^{\mu\nu}=-\frac{\partial^2G}{\partial y_\mu \partial y_\nu},\quad y_\mu=(T,\Omega).
 \end{equation}
Here $G=M-TS-\Omega J$ is the Gibbs free energy with $\Omega=\frac{a}{r_+^2+a^2}$ the angular velocity at the horizon.
In this case, one has
\begin{eqnarray}\label{gK-m}
   g_{W,K}^{\mu\nu} &=&  \left(
           \begin{array}{cc}
             \beta C_\Omega^{\rm K} & \eta^{\rm K}  \\
              \eta^{\rm K}   & \epsilon_T^{\rm K} \\
           \end{array}
         \right),
   \end{eqnarray}
  where $C_\Omega^{\rm K}=T(\partial S/\partial T)_{\Omega}$ is the specific heat at constant angular velocity, $ \eta^{\rm K} =(\partial J/\partial T)_\Omega$ is the off-diagonal element, and $\epsilon_T^{\rm K}=-\frac{3a^4+6a^2r_+^2-r_+^4}{2r_+}=-\frac{\pi(r_+^2+a^2)^2(r_+^2-a^2)}{r_+ C^{\rm K}_J}$ is the isothermal moment of inertia.  The matrix (\ref{gK-m}) could be diagonalized by choosing a new basis of $(dT,d\Omega)\to (dT,d\Omega+ \eta^{\rm K}  dT /\epsilon_T^{\rm K})$ as
\begin{eqnarray}\label{gK-n}
   \tilde{g}_{W,K}^{\mu\nu} &=&  \left(
           \begin{array}{cc}
             \beta C_J^{\rm K} & 0  \\
            0  & \epsilon_T^{\rm K} \\
           \end{array}
         \right)
   \end{eqnarray}
which states that there is always one negative eigenvalue and the GCE is unstable because   $C_J^{\rm K}$ and $\epsilon_T^{\rm K}$ always have opposite signs [see (Left) Fig. 3].

For the KN black hole with $M=1$, its thermodynamic quantities are diverse in GCE. First of all, three specific heats are given in the CE~\cite{Ruppeiner:2008kd} as
\begin{eqnarray}
C_{J,Q}&=&\frac{2\pi K(K^2+L^2+2K)}{L^2-2K}, \nonumber \\
C_{\Omega,Q}&=&\frac{2 \pi K(1+K)^2(K^2+L^2+2K)}{-2K^3-3K^2-2L^2K+2K-3L^2+4}\equiv \frac{(\cdots)}{A} ,\label{hc-KN} \\
C_{J,\Phi}&=&\frac{2\pi K(K^2+L^2+2K)(-L^4-K^2L+KL^2+4L^2+K^3+4K^2+2K-2)}{-K^4+L^2K^3-4K^3-L^2K^2-2K^2+L^4K+2L^2K-4K-2L^4+10L^2-8} \nonumber \\
&\equiv& \frac{(\cdots)}{C} \nonumber
\end{eqnarray}
with $K=\sqrt{1-q^2-a^2}$ and $L=\sqrt{1+a^2}$.
The Davies point is obtained from a condition of $C_{J,Q}\to\infty~(L^2=2K)$ as $a^4+6a^2+4q^2=3$ and the diverging heat capacity $C_{\Omega,Q} \to \infty~(A=0)$ implies
$a^2=\frac{(3-4q^2)q^4}{4(q^2-1)^2}$.  From  $C_{J,\Phi}\to\infty~(C=0)$, one can find its  curve numerically. These all show in (Left) Fig. 4.
All thermodynamic quantities in GCE are computed by using the inverse of  Weinhold metric
\begin{equation}
 g_{W,KN}^{\mu\nu}=-\frac{\partial^2G}{\partial y_\mu \partial y_\nu},\quad y_\mu=(T,\Omega,\Phi)\label{gKN-1n}
 \end{equation}
with  $G=M-TS-\Phi Q-\Omega J$ is the Gibbs free energy.
For $C_{J,Q}$, one finds
\begin{eqnarray}\label{gKN-n}
  g_{W,KN}^{\mu\nu} &=&  \left(
           \begin{array}{ccc}
             \beta C_{\Omega,\Phi}^{\rm KN} & \eta^{\rm KN} &\tilde{\eta}^{\rm KN}   \\
            \eta^{\rm KN}   &  \epsilon_{T,\Omega}^{\rm KN} &\tilde{\epsilon}_{T,\Phi}^{\rm KN} \\
            \tilde{\eta}^{\rm KN} & \tilde{\epsilon}_{T,\Phi}^{\rm KN} & \epsilon_{T,\Phi}^{\rm KN} \\
           \end{array}
         \right),
   \end{eqnarray}
where all components are give by~\cite{Davies:1977bgr}
\begin{eqnarray}
\beta C_{\Omega,\Phi}^{\rm KN}&=&\Big(\frac{\partial S}{\partial T}\Big)_{\Phi,\Omega},\nonumber \\
\eta^{\rm KN}&=& \Big(\frac{\partial S}{\partial \Phi}\Big)_{T,\Omega}=\Big(\frac{\partial Q}{\partial T}\Big)_{\Phi,\Omega}, \nonumber \\
\tilde{\eta}^{\rm KN}&=& \Big(\frac{\partial S}{\partial \Omega}\Big)_{T,\Phi}=\Big(\frac{\partial J}{\partial T}\Big)_{\Phi,\Omega}, \nonumber \\
\epsilon_{T,\Omega}^{\rm KN}&=& \Big(\frac{\partial Q}{\partial \Phi}\Big)_{T,\Omega}, \label{all-the} \\
 \tilde{\epsilon}_{T,\Phi}^{\rm KN}&=& \Big(\frac{\partial Q}{\partial \Omega}\Big)_{T,\Phi}=\Big(\frac{\partial J}{\partial \Phi}\Big)_{T,\Omega}, \nonumber \\
 \epsilon_{T,\Phi}^{\rm KN}&=&\Big(\frac{\partial J}{\partial \Omega}\Big)_{T,\Phi}. \nonumber
\end{eqnarray}
However, it is a formidable task to find all thermodynamic quantities in GCE by diagonalizing the inverse of Weinhold metric
\begin{eqnarray}\label{gKN-d}
  \tilde{g}_{W,KN}^{\mu\nu} &=&  \left(
           \begin{array}{ccc}
             \beta C_{J,Q}^{\rm KN} &0 &0 \\
           0  &  \epsilon_{T,\Omega}^{\rm KN} &0 \\
          0 &0 & \epsilon_{T,\Phi}^{\rm KN}
           \end{array}
         \right).
   \end{eqnarray}
Here, we assume that $C_{J,Q}^{\rm KN}$ and $\epsilon_{T,\Omega}^{\rm KN}(\epsilon_{T,\Phi}^{\rm KN})$ have opposite signs.

Instead, let us  introduce the Ruppeiner metric~\cite{Ruppeiner:2007hr}
\begin{equation}
 g_{R,KN}^{\mu\nu}=-\frac{\partial^2S}{\partial u_\mu \partial u_\nu},\quad u_\mu=(M,J,Q) \label{ru-m}
 \end{equation}
with the entropy $S=\frac{M^2}{8}(2-q^2+2\sqrt{1-q^2-a^2})$.
We note that  Eq.(\ref{ru-m})  is conformal to the Weinhold metric (\ref{gKN-1n}).
In this case, all seven fluctuating cases of $(M,J,Q),(J,Q),(M,Q),\\(M,J), M, J,$ and $Q$ have seven entropy Hessian determinents of $p_1,p_2,p_3, p'_1,p'_2,p''_2,$ and  $p''_1$.
For thermodynamic stability, the entropy Hessian determinants all must   be positive in the whole region of $q^2+a^2\le1$. However, one finds that  $p_3<0$, $p'_1,p'_2$ and $p''_1$ are never negative, while $p_1,p_2,$ and $p''_2$ may have either positive or negative sign depending on the thermodynamic state. This implies that the fluctuation case $(M,J,Q)$ is not stable for  any thermodynamic state, $(J,Q),J,$ and $Q$ are stable for all states, and $(M,J),(M,Q),$ and $M$ may be stable or unstable depending on the thermodynamic state.
The explicit forms of $p_1,p_2,$ and $p''_2$ are given by~\cite{Ruppeiner:2008kd}
\begin{eqnarray}
p_1= \frac{1}{T^2 C_{J,Q}^{\rm KN}}, \quad
p_2= \frac{A}{16K^4}, \quad
p''_2=\frac{C}{16K^4},  \label{ppp}
\end{eqnarray}
where $A$ and $C$ appeared in Eq.(\ref{hc-KN}). $p_1$ has the same sign as $C_{J,Q}^{\rm KN}$ and thus, cyan (pink) regions represent negative (positive) states [see (Left) Fig. 4]. $p_2$ has the same sign as $C_{\Omega,Q}^{\rm KN}$, implying that the left (right) regions  of green curve denote negative (positive) states. Also, $p''_2$ has the same sign as $C_{J,\Phi}^{\rm KN}$, indicating that the lower (upper) regions  of blue curve denote negative (positive) states.
This implies that the whole KN family is thermodynamically unstable in the GCE.

\subsection{Tachyonic instability with scalar coupling}
\begin{figure*}[t!]
   \centering
  \centering
  \includegraphics{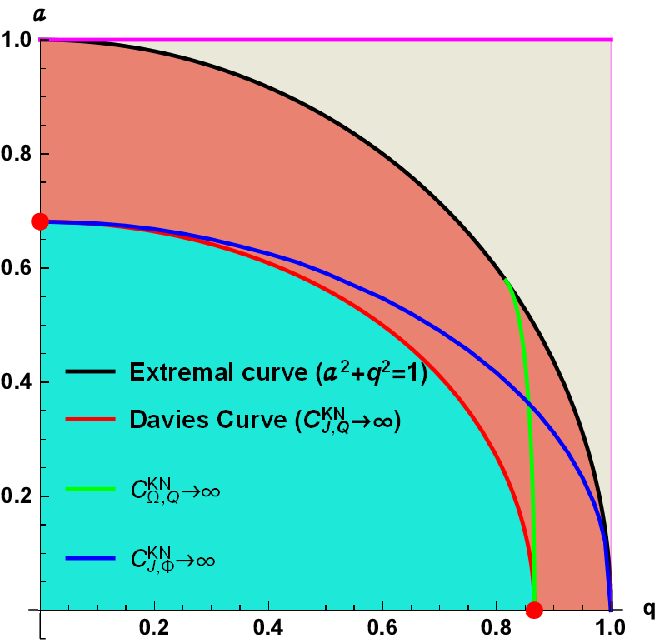}
   \hfill%
  \includegraphics{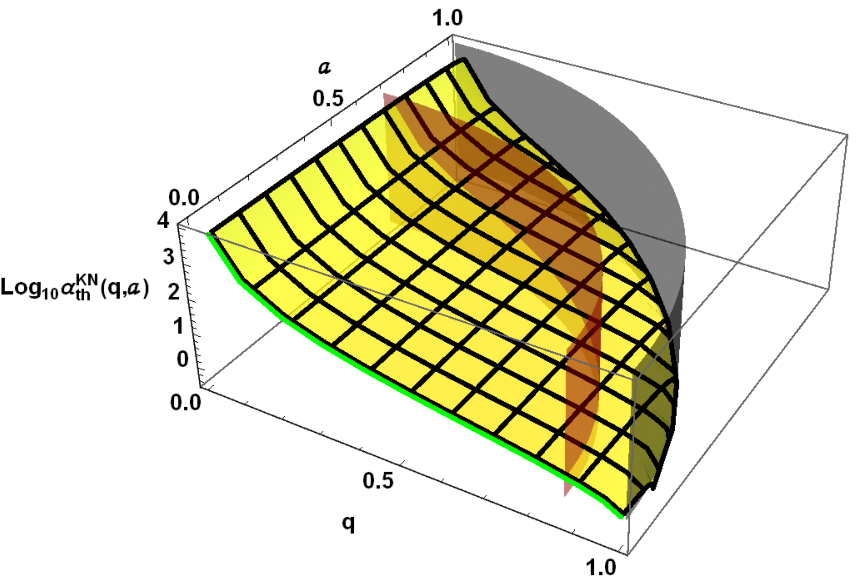}
\caption{(Left) Specific heats   $C^{\rm KN}_{J,Q}(q,a),~C^{\rm KN}_{\Omega,Q}(q,a)$, and $C^{\rm KN}_{J,\Phi}(q,a)$ with $M=1$  for the KN black hole. A red curve represents the diverging specific heat  $C^{\rm KN}_{J,Q}(q,a)$ (Davies curve). Two red dots denote the diverging specific heat $C_Q^{\rm RN}(q)$ at $(q=0.866,a=0)$ and the diverging specific heat $C_J^{\rm K}(a)$ at $(q=0,a=0.681)$, respectively. The cyan (pink) regions indicate $C^{\rm KN}_{J,Q}(q,a)<0$ ($C^{\rm KN}_{J,Q}(q,a)>0$). (Right) A threshold (yellow) surface  of tachyonic instability for KN black hole is $\log_{10} \alpha_{\rm th}^{\rm KN}(q,a)$ when introducing the Maxwell-scalar coupling. A green curve  denotes the threshold curve $\log_{10}\alpha^{\rm RN}_{\rm th}(q)$ [(Right) Fig. 2]  for RN black hole and red (black) boundaries represent the Davies curve and  the extremal KN black hole ($q^2+a^2=1$). For $q\to0$, one could not find its $\log_{10} \alpha_{\rm th}^{\rm KN}(0,a)$ because $\mu^2_{\rm KN} (q,a) \to 0$.}
\end{figure*}

For the Kerr black hole, one has the linearized scalar equation based on Eq.(\ref{egbs})
\begin{equation}
 \Big(\bar{\nabla}^2_{\rm K}-\mu^2_{\rm K}\Big)\delta\phi=0,\label{phi-eq}
\end{equation}
where an effective mass squared is given by
\begin{eqnarray}
&&\mu^2_{\rm K}=-2\alpha \bar{R}^2_{\rm GB}=-\frac{96\alpha M^2(r^6-15r^4 a^2\cos^2\theta+15 r^2a^4\cos^4\theta-a^6\cos^6\theta)}{(r^2+a^2\cos^2\theta)^{6}} \label{KGB}
\end{eqnarray}
in the Boyer-Lindquist coordinates. In the non-rotating limit of $a\to 0$, one recovers $\mu^2_{\rm S}$ (\ref{Smass})  from  $\mu^2_{\rm K}$.
In this case,  we have used the (2+1)-dimensional  hyperboloidal  foliation method (HFM) to derive the threshold curve $\alpha^{\rm K}_{\rm th}(a)$ [see (Right) Fig. 3], which describes  the boundary between stable and unstable Kerr black holes, by considering  $l=0$-scalar mode perturbation~\cite{Zou:2021ybk}. We note that  threshold curve $\alpha^{\rm K}_{\rm th}(a)$ is a decreasing function of $a$ and it has nothing special at the diverging specific heat ($a=0.681$).

To discuss the tachyonic instability of KN black holes in EMS theory (\ref{Action1}), we consider the linearized
perturbation equation
\begin{eqnarray}
\left(\bar{\nabla}_{\rm KN}^2-\mu_{\rm KN}^2\right)\delta\phi=0, \quad
\mu_{\rm KN}^2=\frac{\alpha \bar{F}^2}{2}\label{per-KNeq},
\end{eqnarray}
where an effective mass squared term is given by
\begin{eqnarray}
\mu_{\rm KN}^2=-\frac{\alpha M^2 q^2(r^4-6a^2r^2\cos^2\theta+a^4\cos^4\theta)}{\left(r^2+a^2\cos^2\theta\right)^4}.\label{effmass}
\end{eqnarray}
Under the scalar perturbation with a positive $\alpha>0$, the KN black holes may become unstable in the presence of $\mu^2_{\rm eff}<0$.
In the non-rotating limit of $a\to 0$, one recovers $\mu_{\rm RN}^2$ (\ref{RNmass}) for the RN black hole, whereas one finds $\mu_{\rm KN}^2\to 0$ as $q\to0$.
The latter limit implies  the stability of Kerr black hole for a massless scalar propagation.

After using the HFM~\cite{Lai:2022ppn} to solve the linearized scalar equation numerically, we find from Fig. (Right) Fig. 4 that the threshold (orange) surface [$\log_{10} \alpha_{\rm th}^{\rm KN}(q,a)$]
describes the boundary surface between the stable (lower) and unstable (upper) regions. One finds that there is nothing special when crossing the Davies curve.
Furthermore,  the threshold curve  $\log_{10} \alpha_{\rm th}^{\rm KN}(q,a)$ decreases, as $q\to 1$ and $a \to 1$. This  implies that high charge and high rotation enhance spontaneous scalarization  and   it has nothing to do with a positive heat capacity $C^{\rm KN}_{J,Q}(q,a)>0$ [pink region in (Left) Fig. 4].
\section{SAdS black hole}

\subsection{Thermodynamic instability in GCE}
As was shown for the SAdS black holes~\cite{Prestidge:1999uq}, the negative mode ceases to exist exactly when the specific heat becomes positive.
In this case, the specific heat and moment of inertia  are  given by~\cite{Monteiro:2009tc}
\begin{equation}
C^{\rm SAdS}=2\pi r_+^2\Bigg[\frac{3r_+^2-\frac{3}{\Lambda}}{3r_+^2+\frac{3}{\Lambda}}\Bigg],\quad \epsilon^{\rm SAdS}_T=\frac{r_+^3}{2},
\end{equation}
where  the outer horizon $r_+$  is one real solution to `$ 1-2M/r-\Lambda r^2/3=0$'.
The specific heat blows up at $-\Lambda/3= 0.148/ M^2$, while the moment of inertia is always positive and finite [see (Left) Fig. 5]. The specific heat is negative (positive) for smaller  (larger) SAdS black hole. GCE is unstable for  $-\Lambda/3< 0.148$, while it is stable for $-\Lambda/3> 0.148$.

\begin{figure*}[t!]
   \centering
  \centering
  \includegraphics{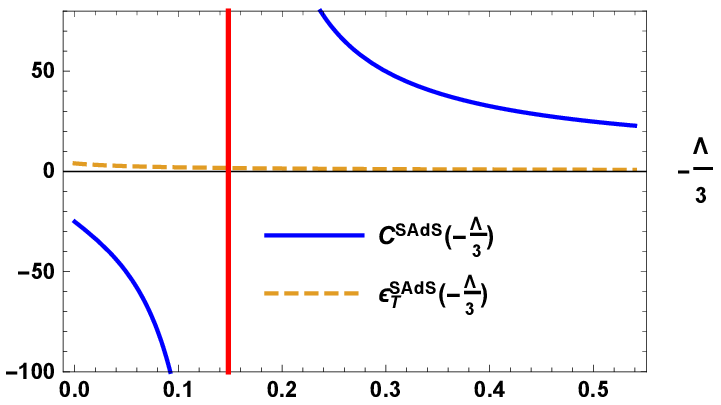}
   \hfill%
  \includegraphics{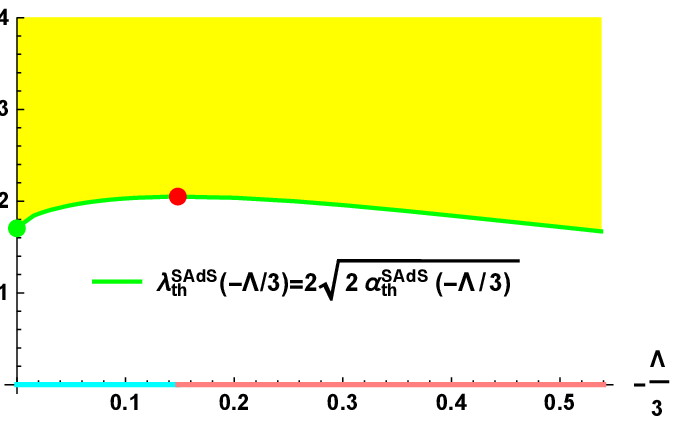}
\caption{(Left) Specific heat  $C^{\rm SAdS}(-\Lambda/3)$ with $M=1$ and isothermal permittivity $\epsilon^{\rm SAdS}_T(-\Lambda/3)$ for SAdS black holes. A red line represents the diverging specific heat at $-\Lambda/3=0.148$. (Right) Threshold curve  of tachyonic instability for SAdS black holes is $\lambda_{\rm th}^{\rm SAdS}(-\Lambda/3)$ when introducing the GB-scalar coupling. A green dot denotes the Schwarzschild black hole at $\lambda_{\rm th}^{\rm S}(0)(=2\sqrt{2\alpha_{\rm th}^{\rm S}})=1.704$.  A red dot represents the diverging specific heat at $-\Lambda/3=0.148$ where  cyan (pink) lines denote negative (positive) specific heats.    }
\end{figure*}

\subsection{Tachyonic instability with scalar coupling}
Here, let us  begin  with the ESGB theory with a negative  cosmological constant $\Lambda<0$~\cite{Doneva:2017bvd}
\begin{equation}
S_{\rm ESGBC}=\frac{1}{16 \pi}\int d^4 x\sqrt{-g}\Big[R-2\Lambda-2(\partial \phi)^2+ \frac{\lambda^2\phi^2}{2} R^2_{\rm GB}\Big],\label{Action1}
\end{equation}
Its linearized scalar equation takes the form
\begin{equation}
 \Big(\bar{\nabla}^2_{\rm SAdS} -\mu^2_{\rm SAdS}\Big) \delta \phi= 0, \label{sads-eq}
 \end{equation}
 where an effective mass squared is given by
 \begin{equation}
 \mu^2_{\rm SAdS}=- \frac{\lambda^2}{4}\bar{R}^2_{\rm GB}.
 \end{equation}
 Taking into account  the separation of variables Eq.(\ref{sep})
and introducing a tortoise coordinate $r_*$ defined by $dr_*=\frac{dr}{1-2M/r-\Lambda r^2/3}$, the radial part of (\ref{sads-eq}) is given by
\begin{equation}
\frac{d^2u}{dr_*^2}+\Big[\omega^2-V_{\rm SAdS}(r)\Big]u(r)=0,
\end{equation}
where the effective potential $V_{\rm SAdS}(r)$  is
\begin{equation} \label{pot-c}
V_{\rm SAdS}(r)=\Big(1-\frac{2M}{r}-\frac{\Lambda r^2}{3}\Big)\Big[\frac{2M}{r^3}+\frac{l(l+1)}{r^2}-\frac{2\Lambda}{3}\Big(1+\lambda^2\Lambda\Big)-\frac{12\lambda^2M^2}{r^6}\Big].
\end{equation}
 We could not obtain the sufficient condition of instability because  $\int^{\infty}_{-\infty}V_{\rm SAdS}(r)dr_*\to \infty$ for any $\Lambda<0$.
To determine the threshold of instability, one  has to solve Eq.(\ref{sads-eq}) with $\omega=i\Omega$ directly,
which may  allow an exponentially growing mode of  $e^{\Omega t}$ as  an unstable mode.
We find the tachyonic  instability for $M=1$ and $-\Lambda/3=0$ as
 \begin{equation}
 \lambda\ge \lambda_{\rm th}^{\rm S}=1.704 \leftrightarrow \alpha \ge \alpha_{\rm th}^{\rm S}=0.363
 \end{equation}
 which corresponds to the tachyonic  instability condition for Schwarzschild black hole. For a threshold curve $\lambda_{\rm th}^{\rm SAdS}(-\Lambda/3)$~\cite{Guo:2020zqm}, see (Right) Fig. 5. It is increasing (like Schwarzschild black hole), approaching the maximum  at the diverging specific heat, and then decreasing.

\section{Discussions}

It has been  confirmed that the negative modes of the Euclidean section for  asymptotically flat black holes reveal the thermodynamic instability of these black holes in the grand canonical ensemble (GCE).
The  Euclidean negative mode with $\lambda_{\rm GPY}$  of the Schwarzschild black hole corresponds to the classical Gregory-Laflamme (GL) instability of the five-dimensional  black string.   When translating the tensor Lichnerowicz equation into the Schr\"{o}dinger-type equation, one found the Zerilli-type potential for $s$-mode which contains a negative region outside the event horizon~\cite{Lu:2017kzi}. This captures the GL instability for obtaining  non-Schwarzschild black hole~\cite{Lu:2015cqa}.  In case of  tachyonic instability for obtaining  asymptotically flat black holes with scalar hair, one usually use the linearized scalar potential  which shows a negative region  near the horizon.   Two instabilities represents the onset for black hole with hairs: GL instability for Ricci-tensor hair and tachyonic instability for scalar hair.
It was shown that the tachyonic instability is regarded as  the hallmark for emerging  scalarized black holes when introducing scalar couplings to the Gauss-Bonnet term or Maxwell term.

In this work,  we have  developed a connection  between thermodynamic instability in GCE and tachyonic instability for asymptotically flat black holes.
All these black holes are thermodynamically unstable in the GCE. On the other hand, the threshold curve $\alpha^{\rm S}_{\rm sc}(M)$ for Schwarzschild black hole is an increasing function of $M$. The threshold curves $\alpha^{\rm RN}_{\rm th}(q)$ [(Right) Fig. 2] for RN black holes and  $\alpha^{\rm K}_{\rm th}(a)$ [(Right) Fig. 3] for Kerr black hole  are decreasing functions
for their negative and positive specific heats.   In the KN black hole, we display  threshold surface  $\log_{10} \alpha_{\rm th}^{\rm KN}(q,a)$ which indicates the boundary surface between stable and unstable regions in (Right) Fig. 4. For all points in the  upper region of  the threshold surface, there are  growing modes and the KN black holes are unstable. The threshold surface  $\log_{10} \alpha_{\rm th}^{\rm KN}(q,a)$ decreases, as $q\to 1$ and $a \to 1$, implying that it has nothing to do with a positive heat capacity $C^{\rm KN}_{J,Q}(q,a)>0$ [pink region in (Left) Fig. 4]. Finally, the threshold curve $\lambda^{\rm SAdS}_{\rm th}(-\Lambda/3)$ [(Right) Fig. 5] for SAdS black holes is increasing for negative specific heat, whereas it is decreasing for its positive specific heat.   This implies that all threshold curves except the Schwarzschild case are decreasing functions even for positive heat capacities.
We conclude  that  tachyonic instability of asymptotically flat black holes which are not related directly  to the their dynamical stability, reflects thermodynamic instability of these black holes in the GCE.

Finally, we mention that there was a discussion on the thermodynamic connection between bald black holes and scalarized black holes in the canonical ensemble~\cite{Santos:2022vet}. In the grand canonical ensemble, the whole KN family is thermodynamically unstable.

 \vspace{2cm}

{\bf Acknowledgments}
 \vspace{1cm}

Y. S. M. was supported by a grant from Inje University for the Research in 2021 (20210040). D. C. Z. acknowledges financial support
from Outstanding Young Teacher Programme from Yangzhou University, No. 137050368.

\end{document}